\documentclass{kapproc} 
\let\footnote\savefootnote
\let\footnotetext\savefootnotetext 
\let\footnoterule\savefootnoterule 
\kluwerbib

\RequirePackage{graphicx}%
\RequirePackage{epsf}%

\newcommand{\psczfig}{
    \begin{figure}
    \vspace{-27pt}
    \begin{center}
    \leavevmode
    \epsfxsize=4.8in    
    \epsfbox{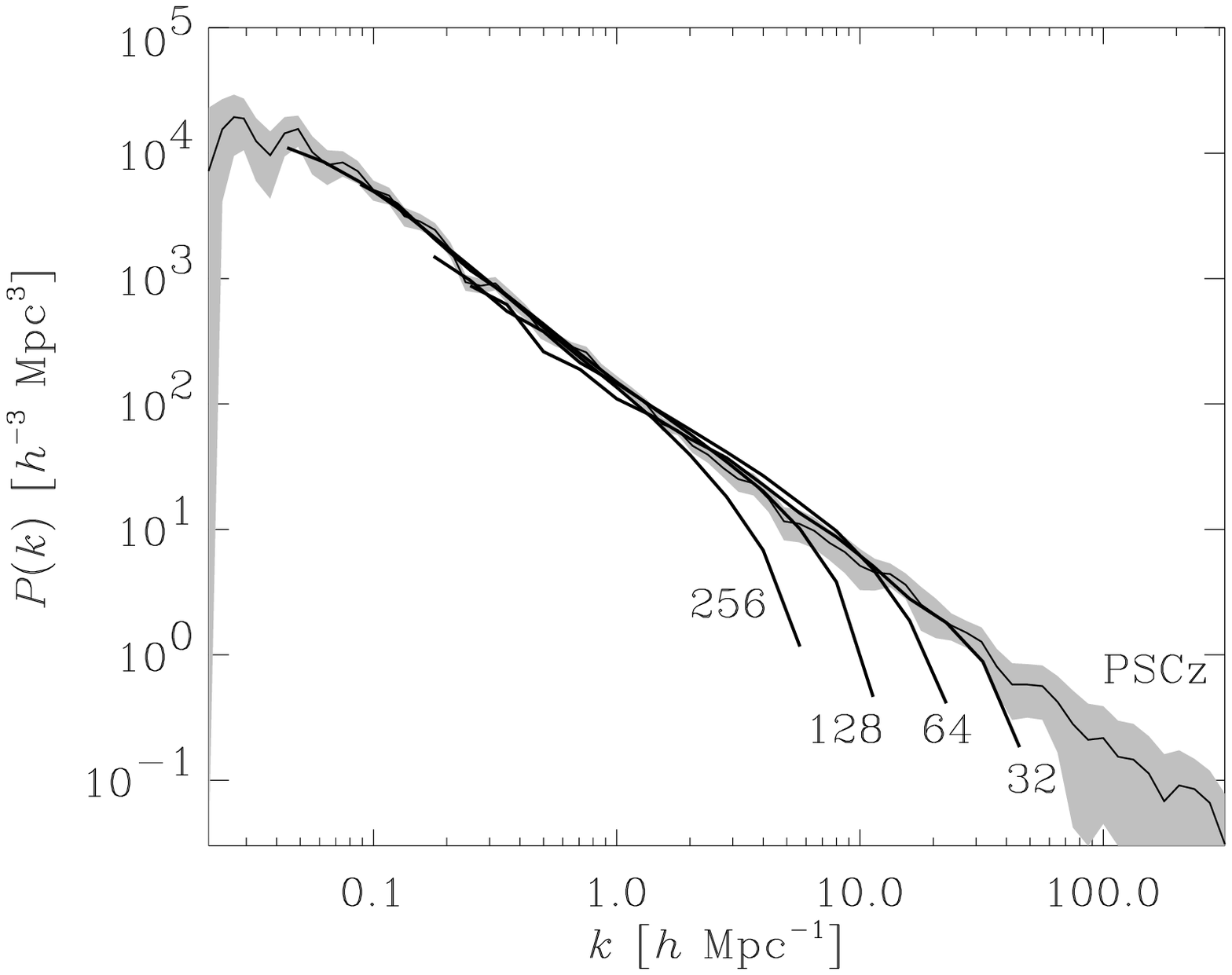}
    \end{center}
      \vspace{-23pt}

    \caption[1]{
      \small The PSCz galaxy power spectrum and its errors, along
      with best fits from the 256, 128, 64, and 32 $h^{-1}\!$ Mpc
      simulations (error bars suppressed).  Each fit is good over a range
      of $\sim$256 in wavenumber.
    \label{sf}
    }
    \end{figure}
}

\begin{document}
  
\articletitle{The PSCz Galaxy Power Spectrum Compared to N-Body
Simulations}

\author{Mark C. Neyrinck, Andrew J.S. Hamilton, Nickolay Y. Gnedin}

\affil{JILA and Dept.\ of Astrophysical \& Planetary Sciences, University of Colorado}
\email{Mark.Neyrinck@colorado.edu, Andrew.Hamilton@colorado.edu, gnedin@casa.colorado.edu}


\begin{abstract}
By comparing the PSCz galaxy power spectrum with haloes from nested and
phased N-body simulations, we try to understand how IRAS infrared-selected
galaxies populate dark-matter haloes.  We pay special attention to the way
we identify haloes in the simulations.

\end{abstract}

The galaxy power spectrum measured by Hamilton \& Tegmark (2002) from the
PSCz redshift survey (Saunders et al, 2000) has a remarkable power-law form
over 4 decades of wavenumber.  Interpreting it in the $\Lambda$CDM paradigm
requires scale-dependent bias between galaxies and dark matter, which
we hoped to reproduce in N-body simulations.

We ran four 256$^3$-particle AP$^3$M dark matter simulations of comoving
boxsize 256, 128, 64, and 32 $h^{-1}\!$ Mpc, each with a fixed Eulerian
softening length of 10 $h^{-1}\!$ kpc.  The boxes are nested in the sense
that the phases of initial fluctuations match in the centers of each box,
ensuring similar structures there. (For a movie depicting this, look at
http:$/\!/$casa.colorado.edu/$\sim$neyrinck/nesthalf.mpg)

The non-trivial step in this project is to go from a collection of dark
matter particles to a set of haloes that could possibly host real galaxies.
We used DENMAX (Gelb \& Bertschinger 1994) with a smoothing length of one
fifth the mean interparticle separation, which gives a mass spectrum
similar to that given by Press-Schechter (1974).  We then varied a low mass
cutoff in the list of haloes to obtain the best fit to PSCz.  However,
there seemed to be an exclusion effect of close pairs because of the rather
large smoothing length, showing itself as a small-scale downturn in the
halo power spectra.

We thus tried ``DENMAX$^2$,'' running DENMAX with half the previous
smoothing length on each halo separately to resolve subhaloes.  We
wanted to characterize sets of DENMAX and DENMAX$^2$ haloes with the same
parameter (mass would no longer do, since the mass returned by DENMAX
changes with smoothing length), so we used central density, which turns out
to be well-correlated with mass in our simulations.

As expected, DENMAX$^2$ resolved more close pairs than DENMAX, extending
the correlation function power law at the small-scale end by a factor of
two or three.  Fig.\ 1 shows the best fits from each of the four nested
simulations.  The central density cutoff in each simulation corresponded to
a mass cutoff of about $10^{11} \hbox{M}_\odot$.

\psczfig

In conclusion, we have succeeded in reproducing the scale-dependent bias of
the PSCz power spectrum at scales $\gtrsim$ 30 $h^{-1}\!$ kpc with N-body
simulations.  Extending the analysis to the smallest scales requires
resolving the substructure of dark matter haloes, which we have done with
an algorithm we call DENMAX$^2$.



\end{document}